\documentclass[conference]{IEEEtran}
\IEEEoverridecommandlockouts
\usepackage{cite}
\usepackage{amsmath,amssymb,amsfonts}
\usepackage{algorithmic}
\usepackage{graphicx}
\usepackage{textcomp}
\usepackage{xcolor}
\def\BibTeX{{\rm B\kern-.05em{\sc i\kern-.025em b}\kern-.08em
    T\kern-.1667em\lower.7ex\hbox{E}\kern-.125emX}}

\usepackage{url}

\usepackage{caption}
\usepackage{subcaption}
\usepackage{xspace}
\usepackage{wasysym}

\usepackage[most]{tcolorbox}
\usepackage{tikz}
\usepackage{fontawesome}
\usetikzlibrary{shapes,decorations,backgrounds}

\usepackage{booktabs}
\usepackage{array}
\usepackage{multirow}
\usepackage{makecell}
\usepackage{amsthm}

\theoremstyle{definition}

\newcommand*{\wiki}[1]{\emph{``#1''}\xspace}

\newcommand{\fakeparagraph}[1]{
  \vspace{6pt}
  \noindent
  \textbf{#1}.
}

\definecolor{boxgrey}{HTML}{F3F3F3}

\begin{document}

\title{Comparing Personalized Relevance Algorithms\\ for Directed Graphs}

\author{\IEEEauthorblockN{Luca Cavalcanti}
\IEEEauthorblockA{\textit{DISI},
\textit{University of Trento}\\
Trento, Italy \\
0009-0001-2529-1639}
\and
\IEEEauthorblockN{Cristian Consonni\textsuperscript{\textsection}}
\IEEEauthorblockA{\textit{Joint Research Centre, European Commission}\\
Ispra, Italy \\
0000-0002-2490-8967}
\and
\IEEEauthorblockN{Martin Brugnara}
\IEEEauthorblockA{\textit{DISI}, \textit{University of Trento}\\
Trento, Italy \\
0000-0002-7348-5147}
\and
\IEEEauthorblockN{David Laniado}
\IEEEauthorblockA{\textit{Eurecat - Centre Tecnol\`ogic de Catalunya}\\
Barcelona, Spain \\
0000-0003-4574-0881}
\and
\IEEEauthorblockN{Alberto Montresor}
\IEEEauthorblockA{\textit{DISI},
\textit{University of Trento}\\
Trento, Italy \\
0000-0001-5820-8216}
}

\maketitle

\begingroup\renewcommand\thefootnote{\textsection}
\footnotetext{Work done while at \textit{Eurecat - Centre Tecnol\`ogic de Catalunya}.}
\endgroup

\begin{abstract}
We present an interactive Web platform that, given a directed graph, allows identifying the most relevant nodes related to a given query node. Besides well-established algorithms such as PageRank and Personalized PageRank, the demo includes Cyclerank, a novel algorithm that addresses some of their limitations by leveraging cyclic paths to compute personalized relevance scores.
Our demo design enables two use cases: (a) algorithm comparison, comparing the results obtained with different algorithms, and (b) dataset comparison, for exploring and gaining insights into a dataset and comparing it with others. 
We provide 50 pre-loaded datasets from Wikipedia, Twitter, and Amazon and seven algorithms. Users can upload new datasets, and new algorithms can be easily added.
By showcasing efficient algorithms to compute relevance scores in directed graphs, our tool helps to uncover hidden relationships within the data, which makes of it a valuable addition to the repertoire of graph analysis algorithms.
\end{abstract}

\begin{IEEEkeywords}
algorithms, directed graphs, personalized pagerank, cyclerank, wikipedia link graph, interactive dashboard
\end{IEEEkeywords}

\section{Introduction}
The relevance of nodes within a graph can be computed globally, or in relation to some specific node. 
The latter is useful when identifying the nodes that are most relevant within a specific context. 
For instance, in the graph of hyperlinks between Wikipedia articles (Wikilink network), that can be viewed as a semantic graph connecting the corresponding concepts and entities~\cite{markusson2016contrasting}, one may want to identify the concepts that are most relevant to a given one. Or in the network of interactions between users in a social media platform, such as Twitter, one can look for the users that are more relevant with respect to a given user. 

One established algorithm from the state of the art for this problem is Personalized PageRank (PPR), a variation of PageRank (PR) in which one can specify one or more nodes as query, and obtain for all the other nodes in the graph a score that captures the degree of relatedness against the nodes of interest. 
This algorithm presents some shortcomings, since it tends to assign a high score to nodes with high global centrality in the graph, regardless of the query node.
For example, in the case of the English Wikipedia, articles such as United States or New York Times (nodes that are globally central) tend to appear among the most relevant topics for any query, even when they are not specifically relevant to the query node.

To overcome these limitations, we have developed an algorithm, called \textsc{Cyclerank}, that leverages the cycles that exist in a directed graph to compute a relevance score for the other nodes in relation to a query node selected by the user~\cite{consonni2020cyclerank}. 

In our web demo, users can compute personalized relevance scores 
on a set of example graphs including Wikilink networks from different language editions, networks of interaction on Twitter about specific topics, and networks of co-purchased products from Amazon. Users can compare the results obtained with different algorithms: Cyclerank, PageRank, Personalized PageRank and several variations that have been introduced in the literature such as CheiRank and 2DRank.

The paper is organized as follows. In Section~\ref{sec:algos}, we describe the algorithms included in our demo, with a focus on CycleRank,  our proposal. Section~\ref{sec:architecture} shows the architecture of our system and its implementation. Section~\ref{sec:usage} compares the results obtained by Cyclerank with those coming from PageRank and various other algorithms, with examples covering a set of diverse topics.
Finally, in Section~\ref{sec:conclusions} we present some directions for future work and we draw our conclusions.

\section{Overview of the Algorithms}\label{sec:algos}

In the following section, we outline briefly the algorithms that are showcased in this demo: \emph{Cyclerank}, \emph{PageRank}, \emph{Personalized PageRank}, \emph{CheiRank} and \emph{2DRank}. 

\fakeparagraph{PageRank}\label{sec:algos:pr}
PageRank~\cite{page1999pagerank} is a metric based on incoming connections, where connections from relevant nodes are given a higher weight. Intuitively, the PageRank score of a node represents the probability that, following a random path in the network, one will reach that node. PageRank can be computed in an iterative process, as the score of a node depends on the score of the nodes that link to it, however more efficient algorithms are available.
The idea at the basis of PageRank is that of simulating a stochastic process in which a user follows random paths in a hyperlink graph. From each node, the algorithm assumes equal probabilities of following any hyperlink included in the page and a certain probability of ``teleporting'' to another random page in the graph. The damping factor $\alpha$, generally assumed to be $0.85$, defines the probability of continuing surfing the graph versus teleporting.

\fakeparagraph{Personalized PageRank}\label{sec:algos:ppr}
Personalized PageRank~\cite{page1999pagerank} is a variant of the original PageRank algorithm, in which teleporting is not directed to all nodes randomly, but to a specific node or set of nodes. In this way, the algorithm models the relevance of nodes around the selected set of reference nodes, as the probability of reaching each of them, when following random walks starting from this chosen set.

\fakeparagraph{Cheirank}\label{sec:algos:cheir}
Chepelianskii~\cite{chepelianskii2010towards} introduced the idea of computing the PageRank score of nodes on the transposed graph and called the algorithm CheiRank. It can be seen as a kind of PageRank based on outgoing instead of incoming connections. 

\fakeparagraph{2DRank}\label{sec:algos:2drank}
Zhirov~\cite{zhirov2010two} combined CheiRank and PageRank to produce a single two-dimensional ranking of Wikipedia articles, 2DRank. This method does not assign a score to each node, but just produces a ranking. 

\fakeparagraph{Cyclerank}\label{sec:algos:cr}
The goal of Cyclerank is to assign to all the nodes in a graph a score that measures their relevance to a given reference node provided as input. Intuitively, a node that is linked from the reference node, but does not link back to it, is likely not to be related to that subject, even if it is relevant in general. Conversely, a node that links to the reference node, but is not linked from it, is likely to be related, but not relevant. Nodes that are linked both from and to a reference node are the ones that we expect to be both relevant and related to it.

We want then to be able to quantify this kind of "mutual relevance" accounting also for the indirect links, i.e., for the number of paths that can be found linking a node from and to the reference node. We do this by counting the number of cycles of various lengths that contain the reference node and any other node. As short distances represent a stronger relationship, short cycles receive a higher weight.

Given a directed graph $G(V,E)$, a reference node $r \in V$ and an integer $K > 1$, the Cyclerank score of any node $i \in V$ with respect to $r$ is given by:
\begin{equation}\label{eq:cr}
CR_{r, K}(i) = \sum_{n=2}^{K}  \sigma(n) \cdot c_{r, n}(i) = \sum_{n=2}^{K} \dfrac{c_{r,n}(i)}{e^n}
\end{equation}
where $c_{r,n}(i)$ is the number of cycles of length $n$ that contain nodes $i$ and $r$, $K$ is a parameter representing the maximum length considered for cycles, and $\sigma(n)$ is the general form of a scoring function that weights the score assigned for each cycle. We set it to be $\sigma=e^{-n}$.

In this way, given a reference node, the Cyclerank score of a node $i$ represents the number of cycles including both the reference node $r$ and node $i$, multiplied by a factor depending on the chosen scoring function and the length of each cycle. For example, for Wikipedia we have experimentally found that the best choice for the scoring function is an exponential damping $\sigma=e^{-n}$.
By definition, the reference node gets the maximum Cyclerank score as it is included in all the cycles considered.

\begin{figure}[t!]
  \centering
  \includegraphics[width=0.75\linewidth]{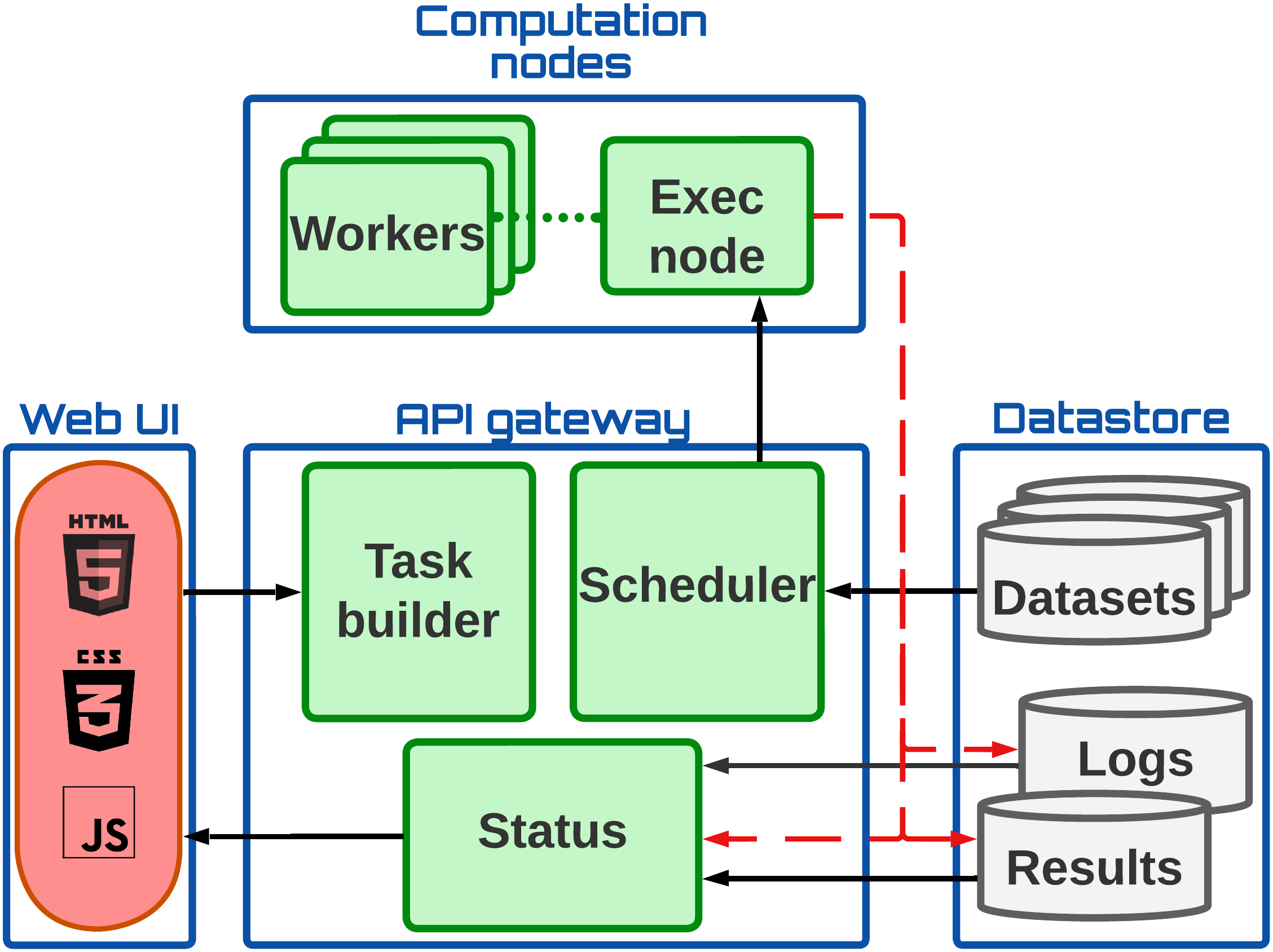}
  \caption{Architecture of the demonstration}
  \label{fig:architecture}
\end{figure}

\section{System Architecture}\label{sec:architecture}

Our system consists of four components, which are presented in the diagram of Figure~\ref{fig:architecture}:
\begin{itemize}
    \item The \emph{Datastore} is responsible for storing and managing datasets. It also provides storage for results and logs produced by the system.
    \item The \emph{API gateway} acts as a mediator between the computational nodes and the web user interface (Web UI). It acts as entry point for all incoming requests from the Web UI and routes them to the relevant computational nodes. Additionally, it could  provide security, authentication, and authorization features for all requests to the system.
    \item The \emph{Computational nodes} are responsible for processing data requests and can be scaled up or down depending on the system's workload. They interact with the data stores to retrieve or store data and then return the results to the API gateway.
    \item The \emph{Web UI} is responsible for collecting user input and presenting the computation results to the user. It interacts with the API gateway to send requests to the computational nodes and display results to the user. The user interface is designed to be user-friendly and responsive to different devices and screen sizes.
\end{itemize}

\noindent Each component is containerized to provide isolation and to simplify development and deployment. Both the API gateway and the computational node (workers) are developed as microservices.

The user interacts with the system through the Web UI. When a request is made the following process is started:
\begin{enumerate}
  \item a task, which is a triple consisting of a dataset, an algorithm and a set of parameters is built by the Task Builder component and is sent to the Scheduler;
  \item when the Scheduler receives the task, it fetches the dataset and invokes an Executor node;
  \item the computation needed to perform the task is off-loaded to the worker nodes; while the computation is running, the Status component polls the Executor node to monitor its progress;
  \item when the computation is completed, results and logs are written to the datastore; the Status component can access them in response to user requests;
  \item finally, the API returns the results of the completed task to the Web UI which displays them to the user.
\end{enumerate}

Our demo design enables the possibility of adding new algorithms to the demo. We do not offer this possibility directly because it would expose us to several security issues that are outside the scope of this demo.

\section{Results and Discussion}\label{sec:results}

\begin{table*}[ht!]
  \centering
  \small
  \caption{Top-5 articles with the highest PR ($\alpha = 0.85$), CR ($K=3, \sigma=e^{-n}$) and PPR ($\alpha = 0.3$) scores computed on the 2018-03-01 English Wikipedia snapshot. The reference articles for CR and PPR are \wiki{Freddie Mercury,} and \wiki{Pasta}.}
  \resizebox{.95\textwidth}{!}{
\begin{tabular}{@{}rp{3.30cm}p{3.30cm}p{3.30cm}p{3.30cm}p{3.30cm}@{}}
  \toprule
    \textbf{page}
  & %
  & \multicolumn{2}{c}{\thead{\bfseries Freddy Mercury}}%
  & \multicolumn{2}{c}{\thead{\bfseries Pasta}}%
  \\ %
  \cmidrule(l){2-2}%
  \cmidrule(l){3-4}%
  \cmidrule(l){5-6}%
    \textbf{\#}%
  & \multicolumn{1}{c}{\thead{\bfseries PageRank}}  %
  & \multicolumn{1}{c}{\thead{\bfseries Cyclerank}} & \multicolumn{1}{c}{\thead{\bfseries Pers.PageRank}}%
  & \multicolumn{1}{c}{\thead{\bfseries Cyclerank}} & \multicolumn{1}{c}{\thead{\bfseries Pers.PageRank}}%
  \\%
  1  & United States                                                                                     %
     & Freddie Mercury                              & Freddie Mercury                                    %
     & Pasta                                        & Pasta                                              %
     \\%
  2  & Animal                                                                                            %
     & Queen (band)                                 & Queen (band)                                       %
     & Italian cuisine                              & Bolognese sauce                                    %
     \\%
  3  & Arthropod                                                                                         %
     & Brian May                                    & The FM Tribute Concert                             %
     & Italy                                        & Carbonara                                          %
     \\%
  4  & Association football                                                                              %
     & Roger Taylor                                 & HIV/AIDS                                           %
     & Spaghetti                                    & Durum                                              %
     \\%
  5  & Insect                                                                                            %
     & John Deacon                                  & Queen II                                           %
     & Flour                                        & Italy                                              %
     \\%
  \bottomrule
\end{tabular}%
}

  \label{table:comparison_wikipedia_pr-cr-ppr}
\end{table*}

\begin{table*}[ht!]
  \centering
  \small
  \caption{Top-5 articles with the highest PR ($\alpha = 0.85$), CR ($K=5, \sigma=e^{-n}$), and PPR ($\alpha = 0.85$) scores computed on the Amazon co-purchase dataset. The reference items for CR and PPR are \wiki{1984,} and \wiki{The Fellowship of the Ring.}}
  \resizebox{.95\textwidth}{!}{
\begin{tabular}{@{}rp{3.30cm}p{3.30cm}p{3.30cm}p{3.30cm}p{3.30cm}@{}}
  \toprule
    \textbf{item}
  & %
  & \multicolumn{2}{c}{\thead{\bfseries 1984}}%
  & \multicolumn{2}{c}{\thead{\bfseries The Fellowship of the Ring}}%
  \\ %
  \cmidrule(l){2-2}%
  \cmidrule(l){3-4}%
  \cmidrule(l){5-6}%
    \textbf{\#}%
  & \multicolumn{1}{c}{\thead{\bfseries PageRank}} %
  & \multicolumn{1}{c}{\thead{\bfseries Cyclerank}} & \multicolumn{1}{c}{\thead{\bfseries Personalized PageRank}} %
  & \multicolumn{1}{c}{\thead{\bfseries Cyclerank}} & \multicolumn{1}{c}{\thead{\bfseries Personalized PageRank}} %
  \\%
  1  & Good to Great                                                                                             %
     & Animal Farm                                  & The Catcher in the Rye                                     %
     & The Hobbit                                   & The Silmarillion                                           %
     \\%
  2  & The Catcher in the Rye                                                                                    %
     & Fahrenheit 451                               & Lord of the Flies                                          %
     & The Return of the King                       & The Hobbit                                                 %
     \\%
  3  & DSM-IV                                                                                                    %
     & The Catcher in the Rye                       & Animal Farm                                                %
     & The Silmarillion                             & Harry Potter (Book 1)                                      %
     \\%
  4  & The Great Gatsby                                                                                          %
     & Brave New World                              & Fahrenheit 451                                             %
     & The Two Towers                               & Harry Potter (Book 2)                                      %
     \\%
  5  & Lord of the Flies                                                                                         %
     & Lord of the Flies                            & To Kill a Mockingbird                                      %
     & Unfinished Tales                             & The Return of the King                                     %
     \\%
  \bottomrule
\end{tabular}%
}

  \label{table:comparison_amazon_pr-cr-ppr}
\end{table*}

\begin{table*}[ht!]
  \centering
  \small
  \caption{Top-5 articles with the highest Cyclerank ($K=3, \sigma=e^{-n}$) scores computed on different Wikipedia language editions (\texttt{de}, \texttt{es}, \texttt{fr}, \texttt{it}, \texttt{nl}, \texttt{pl}) using the reference article \wiki{Fake news}.}
  \resizebox{.975\textwidth}{!}{
\begin{tabular}{@{}rp{2.75cm}p{2.75cm}p{2.75cm}p{2.75cm}p{2.75cm}p{2.75cm}@{}}
  \toprule
    \textbf{\#}%
  & \multicolumn{1}{c}{\thead{\bfseries Fake News (de)}} %
  & \multicolumn{1}{c}{\thead{\bfseries Fake news (en)}} %
  & \multicolumn{1}{c}{\thead{\bfseries Fake news (fr)}} %
  & \multicolumn{1}{c}{\thead{\bfseries Fake news (it)}} %
  & \multicolumn{1}{c}{\thead{\bfseries Nepnieuws (nl)}} %
  & \multicolumn{1}{c}{\thead{\bfseries Fake news (pl)}} %
  \\%
  \cmidrule(l){2-2}%
  \cmidrule(l){3-3}%
  \cmidrule(l){4-4}%
  \cmidrule(l){5-5}%
  \cmidrule(l){6-6}%
  \cmidrule(l){7-7}%
  1  & Barack Obama                                 %
     & CNN                                          %
     & Ère post-vérité                              %
     & Disinformazione                              %
     & Facebook                                     %
     & Dezinformacja                                %
     \\%
  2  & Tagesschau.de                                %
     & Facebook                                     %
     & Donald Trump                                 %
     & Post-verità                                  %
     & Journalistiek                                %
     & Propaganda                                   %
     \\%
  3  & Desinformation                               %
     & US pres. election, 2016                      %
     & Facebook                                     %
     & Bufala                                       %
     & Hoax                                         %
     & Media społecznościowe                        %
     \\%
  4  & Fake                                         %
     & Propaganda                                   %
     & Hoax                                         %
     & Debunker                                     %
     & Donald Trump                                 %
     & -                                            %
     \\%
  5  & Donald Trump                                 %
     & Social media                                 %
     & Alex Jones (complotiste)                     %
     & Clickbait                                    %
     & -                                            %
     & -                                            %
     \\%
  \bottomrule
\end{tabular}%
}

  \label{table:comparison_wikipedia_multilang}
\end{table*}

In this section we present the demo resources and interface, and how to build queries. Finally, we showcase some use cases and discuss them through concrete examples.

\subsection{Resources}

The demo can be accessed at \url{https://cricca.disi.unitn.it/cyclerank-demo/}, the code is available on GitHub at \url{https://github.com/CycleRank/cyclerank-demo}.

\subsection{Datasets}

In principle, all algorithms showcased can be run on any directed graph. We provide these datasets because they present interesting use cases for the problem of personalized relevance in graphs:
\begin{itemize}
  \item \emph{WikiLinkGraphs}~\cite{consonni2019wikilinkgraphs}:  networks of links between Wikipedia articles (wikilinks). Yearly snapshots for 9 languages (\texttt{de}, \texttt{en}, \texttt{es}, \texttt{fr}, \texttt{it}, \texttt{nl}, \texttt{pl}, \texttt{ru}, and \texttt{sv}) for the years 2018, 2013, 2008, and 2003.
  \item \emph{Amazon}~\cite{leskovec2007dynamics}: product ids and metadata about $548,552$ different products (Books, music CDs, DVDs) that were co-purchased (``Customers who bought X also bought Y'') on Amazon. 
  \item{Twitter}: networks of users that made some tweet containing a set of keywords of interest. Each user is a node and two users are connected if they have interacted in some way (retweet, reply, quote or mention). In particular we provide two datasets from Twitter: \texttt{cop27} (COP27 climate conference); and \texttt{8m} (8th of March, International Women's Day).
\end{itemize}
Users can upload their own dataset to run the provided algorithms on their dataset. We support commonly used graph formats such as: \texttt{edgelist (CSV)}\footnote{A comma-separated value with source and target indexes, see: \url{https://gephi.org/users/supported-graph-formats/csv-format/}.}, \texttt{pajek}\footnote{A format where each line is an element, and the list of edges follows the list of nodes, see \url{https://gephi.org/users/supported-graph-formats/pajek-net-format/}.}, and our own \texttt{ASD} format. We detail the supported formats in the Instructions page of the demo.

\subsection{Usage}\label{sec:usage}

The Web UI enables users to select the dataset and the algorithm to execute, along with its parameters. For Cyclerank, these parameters include a reference node $r$, the maximum cycle length $K$, and the scoring function $\sigma$, as defined in Equation~\ref{eq:cr}. For PageRank, CheiRank and 2DRank, it is sufficient to specify the value of the transition probability $\alpha$, while their personalized  variants also take a reference node $r$.

\begin{figure}[ht!]
    \centering
    \begin{minipage}[t]{0.95\columnwidth}%
\centering
\tikzstyle{background rectangle}=[thin,draw=black]
\begin{tikzpicture}[show background rectangle,rounded corners=5pt]

\node[align=left, text width=0.95\columnwidth]{

\vspace{0.5em}
\footnotesize{Comparison id: \texttt{3a73ff34-8720-4ce8-859e-34e70f339907}} 
\hspace*{\fill} \tiny{\faTrashO} \; \\
\vspace{1em}

\resizebox{1.0\textwidth}{!}{
  \bgroup
  \def\arraystretch{1.25}%
  \begin{tabular}{llllll}
    \cmidrule{3-6}
      \textbf{Id}           %
    &                       %
    & \textbf{Dataset}      %
    & \textbf{Algorithm}    %
    & \textbf{Source}       %
    & \textbf{Parameters}   %
    \\%
    \cmidrule(l){3-6}
    0 & \makecell{\XBox} & \texttt{enwiki 2018-03-01} & Cyclerank      & \texttt{Fake news} & $k=3, \sigma=\exp$ \\
    1 & \makecell{\XBox} & \texttt{enwiki 2018-03-01} & PageRank       & \texttt{-}         & $\alpha = 0.3$     \\
    2 & \makecell{\XBox} & \texttt{enwiki 2018-03-01} & Pers. PageRank & \texttt{Fake news} & $\alpha = 0.3$     \\
    \bottomrule
  \end{tabular}
  \egroup
}

};

\node[xshift=3ex, yshift=-0.75ex, overlay, fill=white, draw=white, above 
right] at (current bounding box.north west) {
\textit{Query Set}
};

\end{tikzpicture} 
\end{minipage}
    \caption{Task builder interface}\label{fig:task_builder}
\end{figure}

\noindent The user can create arbitrary query sets by choosing different reference nodes, datasets and parameters, which are then visualized in the task builder interface, as presented in Figure~\ref{fig:task_builder}. A query set defines a task, and a unique identifier is assigned to it, serving as a permalink to retrieve its results.
A query set can be emptied by clicking on the trash bin symbol (\faTrashO) on the top right, or queries can be eliminated individually (\XBox).
 
\subsection{Use cases}
We present here two use cases for our demonstration: algorithm comparison and dataset comparison.

\fakeparagraph{Algorithm comparison} Our demo can showcase the results of all the presented algorithms allowing users to compare the strenghts and limitations of each algorithm.
Tables~\ref{table:comparison_wikipedia_pr-cr-ppr} and~\ref{table:comparison_amazon_pr-cr-ppr} present a comparison between the top-5 results with the highest scores obtained by PageRank, Cyclerank and Personalized PageRank, computed with different parameters on the wikilink graph of the English Wikipedia taken as of March 1st, 2018 and the Amazon co-purchase graph, respectively. 

These results highlight the limitation of Personalized PageRank: in Table~\ref{table:comparison_wikipedia_pr-cr-ppr}, Personalized PageRank promotes results that have a very high in-degree and have among the highest values of the PageRank score in the overall network. This problem has been studied in depth in~\cite{consonni2020cyclerank}. Similarly, on the Amazon co-purchase graph in Table~\ref{table:comparison_amazon_pr-cr-ppr}, Personalized PageRank suggest popular items such as the \emph{``Harry Potter''} book series, while Cyclerank does not.

\fakeparagraph{Dataset comparison} Another use case for our demo consists in applying the same algorithm on different datasets to gain new insights. Table~\ref{table:comparison_wikipedia_multilang} presents a cross-cultural comparison between the relevant nodes in the Wikipedia link graph identified by Cyclerank starting from the article \wiki{Fake news} for different editions, showing how the same concept is framed differently across different language communities. A similar analysis can also be performed by comparing snapshots of a graph at different points in time, another functionality available in the demo.

\section{Conclusions and Future Work}\label{sec:conclusions}
In this demonstration, we showcased Cyclerank, a novel algorithm that leverages cyclic paths to identify the most relevant nodes related to a given query node. We also described two use cases:
\begin{itemize}
  \item \emph{algorithm comparison}: we compared Cyclerank with $6$ established algorithms: PageRank, CheiRank, 2DRank and their personalized variants. We highlighted how, on different datasets from Wikipedia and Amazon, Cyclerank could produce more relevant answers and mitigate the issue of over-representation of popular nodes. Our demo design enables the possibility of adding new algorithms to the demo.
  \item \emph{dataset comparison}: we showed how Cyclerank can be used to compare different datasets, providing insights into the relationships within the data. We presented a cross-cultural comparison of the results obtained on the Wikipedia link graph 
  for the same topic in six different languages. Users can also upload their own datasets. We support three dataset formats and we plan to add new ones in the future.
\end{itemize}
\noindent In summary, our flexible demo design enables the comparison of relevance ranking algorithms on directed graphs.

\section*{Acknowledgment}

We would like to thank Joan Massachs and Juli\'an Vicens for their help with collecting the Twitter datasets. This work was supported by the REMISS project with grant PLEC2021-007850 funded by MCIN/AEI/10.13039/501100011033, and by the Horizon Europe CLOUDSTARS project (101086248).

\bibliographystyle{ieeetr}
\bibliography{biblio}

\end{document}